
\overfullrule=0pt
\pubnum={92-13}
\date={ }
\titlepage
\hoffset=-0.2in
\hsize=6.9in
\voffset=-0.2in
\vsize=9.3in

\def\del{\delta}
\def\F{\widetilde{F}}

\bigskip

\title{Salient Features of High-Energy Multiparticle Distributions\break
Learned from Exact Solutions of 1-d Ising Model}

\bigskip

\author{Ling-Lie Chau$^*$  and Ding-Wei Huang$^{\dagger}$}

\address{\it $^{*\dagger}$ Department of Physics, University of California,
Davis, CA~~95616;\break
$^*$ CNLS, Los Alamos National Laboratory, Los Alamos, NM 87545.}

\vfill

\abstract

We have derived explicit expressions in the 1-d Ising model
for multiplicity distributions $P_{\del\xi}(n)$ and factorial moments
$F_q(\del\xi)$.
We identify the salient features of
$P_{\del\xi}(n)$ that lead to
scaling, $F_q(\del\xi)=\F_q[F_2]$, and universality.
These results compare well with the presently available high-energy data
of $\bar{p}p$ and $e^+e^-$ reactions.
We point out the important features that
should be studied in future higher-energy experiments of multiparticle
productions in $pp$, $\bar{p}p$,
$ep$, $e^+e^-$, and $NN$.
We also make comments on comparisons with KNO and negative-binomial
distributions.

\vfill
\endpage

 \advance \abovedisplayskip by -0.15in
 \advance \belowdisplayskip by -0.15in
 \advance \abovedisplayshortskip by -0.15in
 \advance \belowdisplayshortskip by -0.15in

\noindent{\bf Introduction and Summary}

Multiparticle productions in hadronic interactions or in jet-hadronizations
are effectively one-dimensional (1-d) distributions in the rapidity variable
or its equivalence due to the sharp cut-off in transverse momenta;
thus we can study them in 1-d models.
With the two interactions, nearest-neighbor influence and an external
agitation, 1-d Ising provides the simplest model in which
multiparticle distributions can be calculated.
The reasoning of our using Ising model for multiparticle is very similar to
that of using Ising model for lattice gas, except the space now
in multiparticle production is rapidity.  Ising model is not
a dynamical model that governs all detail dynamics of multiparticle
productions,
as it is not for phase transitions;
however it may capture the important
underlying laws in the seemingly complicated multiparticle productions
as it has for phase transitions.
Indeed, we show here that we have learned
the salient mathematical structures in $P_{\del\xi}(n)$
that lead to scaling
and universality, as well as found a framework to analyze data of high-energy
multiparticle productions in
$pp$, $\bar{p}p$, $ep$, $e^+e^-$, and $NN$.
($\del\xi$ in $P_{\del\xi}(n)$ is the interval of the 1-d rapidity variable;
its full-range will be denoted by $\Delta\xi$.)

As reported in a previous publication\attach{[1]},we found that the Ising model
naturally gives scaling $F_q(\del\xi)=\F_q[F_2]$, $q=3,4,5,\cdots$ and
universality,
which states
that all dynamical dependences are contained in $F_2(\del\xi)$, thus
$\F_q[F_2]$ is an universal function for different reactions at different
energies.
In our previous paper\attach{[1]} we had derived explicit expressions for
$F_q(\del\xi)$ and $\F_q[F_2]$.
Here we perform a different calculation in the 1-d  Ising model and
give explicit expressions for multiplicity distributions
$P_{\del\xi}(n)$,
which provides the most basic information
on the multiparticle distributions.
Other distributions, like factorial moments $F_q(\del\xi)$ and
their scaling and universality $F_q(\del\xi)=\F_q[F_2]$, can be derived from
it.
Our $P_{\del\xi}(n)$ is expressed as a $\gamma^n$-weighted cluster $l$-sum of
$n$ particles $(c^l/l!) C^{n-1}_{l-1}$,($C^{n-1}_{l-1}$ is the binomial);
$\gamma$ and $c$ contain the dynamical parameters of the theory
and they are determined by experimental values of $\VEV{n}_{\del\xi}$
and $\VEV{n^2}_{\del\xi}$;
interestingly that we can also express $P_{\del\xi}(n)$ as a
$\gamma^n$-weighted
sum of products of a Poisson and a negative binomial; see Eq.(4) and the
paragraph followed.

After fixing the two parameters of the model by fitting with data
$\VEV{n}_{\del\xi}$ and
$\VEV{n^2}_{\del\xi}$ at various values of $\del\xi$,
the multiparticle distributions $P_{\del\xi}(n)$ as functions of $n$ are
determined in the
model and constitute as predictions of the model.  They compare well
with data, better for larger region of $\del\xi$, Figs.(1a) and (1b).
Interestingly the two dynamical parameters in the model become rather
flat in their dependences on $\del\xi$ for larger intervals of  $\del\xi \geq
3$.
This implies that in the larger $\del\xi$-intervals Ising model with
$\del\xi$-independent
interaction-parameters ($A_{\del\xi}$ and $B_{\del\xi}$) can also describe the
$\del\xi$-dependences
in $P_{\del\xi}(n)$ and $F_q(\del\xi)$;
thus future data at higher energies, which make larger $\del\xi$-intervals
available,
will provide further important information.

{}From $P_{\del\xi}(n)$, we can calculate
factorial moments
$$
F_q(\delta \xi) \equiv   [\VEV{n(n-1)\ldots (n-q+1)}_{\delta \xi}] ~
[\VEV{n}_{\delta \xi}]^{-q},~~~q=1,2,3,\cdots , \eqno(1)
$$
where $\VEV{(~)}_{\delta \xi}\equiv
\Sigma_{n=0}^{\infty}~\{(~)~P_{\del\xi}(n)\}$~.
These factorial moments $F_q$ of
multiparticle-productions were pointed out to be important to study in Ref.[2]
and have been known in Ref.~[3] to have the scaling
behavior $F_q(\del\xi)=\F_q[F_2]$, and known in Ref.[4] to have
universality.
The data considered include
$\bar p p(\sqrt s=200$ to 900 GeV)\attach{[5]},
quark jets from $e^+e^- (\sqrt s=91$ GeV)\attach{[6]}, as well as various
nuclear
reactions $(p, ^{16}O, ^{32}S$ at 200 GeV/nucleon).\attach{[7]}
In Ref.[1] we showed that 1-d Ising model can produce these salient features.
Here we compare these data with our parameter-free universal
distributions $\F_q[F_2]$ predicted from 1-d Ising model,
see Figs.(2), (3a,b), (4a,b) and (5a,b).
The agreement is excellent in the smaller $F_2(\del\xi)$ region
which comes from larger $\del\xi$.
We anticipate that this trend will continue at higher-energies where the
available $\del\xi$-interval will increase.

Our Ising results have no KNO scaling.\attach{[8]}
Single-negative-binomial-distributions (NB)\attach{[9,10,11]} have been popular
in characterizing
multiparticle distributions, but their dynamical origin is not clear.
Ising distributions $P_{\del\xi}(n)$ are explicitly different functions from
single-negative-binomial distributions, though numerically they are similar
in comparing with current data, see Fig.(1c).

It is suprising, yet satisfying,
that a simple model like Ising captures
so many essential features of multiparticle distributions.
Later toward the end of the paper we shall describe more specifically the
important features
to be studied in future higher-energy experiments.

\noindent{\bf Multiplicity Distributions from 1-d Ising Model}

First we give a brief description of the 1-d Ising calculation.
The Ising-model Hamiltonian is
$H=-\varepsilon \Sigma^N_{(i,j)} S_i \cdot S_j - b
\Sigma^N_i S_i$, where $(i,j)$
means $j=i\pm 1$ and $i$ is summed over all lattice sites $N$; $\varepsilon$
signifies the strength of next-neighbor influence
and $b$ the strength of an agitating field.
Conventionally $S_i$ represents  the spin values $\pm{1 \over 2}$ at site $i$.
Same as for lattice gas, defining $n_i\equiv({1\over 2}+S_i)=\{^1_0$ we can
interpret $n_i$ as one or no particle
production at site $i$.
In this interpretation, $b$ represents the agitation that produces particles
and $\varepsilon$ the short-range next-neighbor influence
among the particles.

The multiplicity distribution in a sublattice $N/M$ is
$
P_M(n)=(1/Z)\{\{\Sigma_{\{n_i\vert\Sigma n_i=n\}}e^{-\beta H}\}\}
$,
where
$\{n_i\vert \Sigma n_i=n\}$ means summing over all possible configurations
of $n_i$ in the sublattice $N/M$ with
the constraint $\Sigma n_i=n$.
Note that $\Sigma_{n=0}^{N/M} P_M(n) =1$ and $Z=\Sigma_{n=0}^{N/M} \{\{~
\}\}$-part of $P_M(n)$.
After some calculations,
we obtain
$$
P_M(n)=(1/z_M) \{~ \gamma^n~
 \Sigma_{l=1}^n~
[~(N/M)~C_{l-1}^{{(N/M)}-n-1}~e^{-4\varepsilon\beta l}~l^{-1}~]~C_{l-1}^{n-1}~
{}~\} ,\eqno(2)
$$
for $n\ne 0$; and for $n=0$, $P_M(0)=1/z_M$.  In Eq.(2)
$l$ is the number of clusters among $n$;
$\gamma=e^{2\beta b}$; $z_M=
[\Sigma_{n=0}^{N/M}~\{~\}$-part of Eq.(2)]
$=\{(1+\gamma)/2 +[(1-\gamma)^4/4+\gamma
e^{-4\beta\varepsilon}]^{1/2}\}^{N/M}$
$ +
\{(1+\gamma)/2 -[(1-\gamma)^4/4+\gamma
e^{-4\beta\varepsilon}]^{1/2}\}^{N/M}
$ the normalization factor, which is related to $Z$ by
$z_M=e^{-\beta(\varepsilon-b)N/M}~Z$; and $C_k^j=j![k!(j-k)!]^{-1}$ the
binomial.
Due to the $\gamma^n$-dependence of $\{~\}$-part of $P_M(n)$ in Eq.(2), we can
calculate
$\VEV{n}_M$ and $\VEV{n^2}_M$ by $\partial \gamma$-differentiating the
explicitly-given $z_M$:
$\VEV{n}_M=\Sigma_{n=0}^{N/M} nP_M(n)
=(1/z_M)\partial z_M/\partial \gamma$
and $\VEV{n^2}_M=\Sigma_{n=0}^{N/M} n^2P_M(n)
=(1/z_M)\partial^2 z_M/(\partial \gamma )^2
 +(1/z_M)\partial z_M/\partial \gamma$,
which are functions of $\beta\varepsilon,
\beta b$ and $N$:
$\langle n\rangle_M$
$=(N/2M)+(N/2M)cos(2\phi)$ $(1-a^{N/M})$ $(1+a^{N/M})^{-1}$
and
$\langle n^2\rangle_M$
$=\langle n\rangle_M^2+(N/M)^2cos^2(2\phi) a^{N/M}$
$(1+a^{N/M})^{-2}$ $+(N/4M)sin^2(2\phi)$ $(1+a)(1-a)^{-1}$
$(1-a^{N/M})$ $(1+a^{N/M})^{-1}$
with
$2\phi \equiv$
$tan^{-1}[e^{-2\beta\epsilon} ~sinh^{-1}(\beta b)]$
and
$a\equiv$
$\{cosh(\beta b) - [sinh^2(\beta b) +e^{-4\beta\epsilon}]^{1/2}\}$
$\{cosh(\beta b) + [sinh^2(\beta b) +e^{-4\beta\epsilon}]^{1/2}\}^{-1}$.

To make the connection with multiparticle productions,
we take the continum limit $N\rightarrow\infty$ with
fixed $\langle n\rangle_{\del\xi}$ and $\langle n^2\rangle_{\del\xi}$,
which result in two finite-parameters
$A_{\del\xi}=N e^{(-4\beta\varepsilon)}$ and $B_{\del\xi}=-\beta b$,
(notice that $a<1$ and all $a^{N/M}\rightarrow 0$).\attach{[1]}
Further we interprete\attach{[12]} $M=\Delta\xi/\del\xi$, $\Delta\xi$ being
the whole range of rapidity available, and obtain
$$
\VEV{n}_{\del\xi}=A_{\del\xi}~ [2 sinh(B_{\del\xi})]^{-2}~
(\del\xi/\Delta\xi) ~~~~,~~~~
\VEV{n^2}_{\del\xi}=\VEV{n}^2_{\del\xi} +\VEV{n}_{\del\xi}~
coth(B_{\del\xi})  ~~;\eqno(3)
$$ and
$$\eqalignno{
P_{\del\xi}(n)&=(1/z_p) ~\{~\gamma^n~
\Sigma_{l=1}^n~ [~c^l~/~ l!~] ~C^{n-1}_{l-1}~\}  ~~;
{}~~~~~~~P_{\del\xi}(0)=1/z_p~~,&(4)\cr
z_p&=exp[\gamma c/(1-\gamma)]   ~~,&(5)
}$$
where $z_p=z_M$ at $N\rightarrow\infty$
and $z_p=\Sigma_{n=0}^{\infty}~\{~\}$-part of Eq.(4);
using Eq.(3), $\gamma$ and $c$ can be expressed in terms of $\VEV{n}_{\del\xi}$
and $\VEV{n^2}_{\del\xi}$:
$
\gamma = e^{2\beta b} = e^{-2B_{\del\xi}} =
[\VEV{n^2}_{\del\xi}-\VEV{n}_{\del\xi}^2-\VEV{n}_{\del\xi}]
[\VEV{n^2}_{\del\xi}-\VEV{n}_{\del\xi}^2+\VEV{n}_{\del\xi}]^{-1}
$ and
$
c = A_{\del\xi}~(\del\xi/\Delta\xi) =
4\VEV{n}_{\del\xi}^3
[\VEV{n^2}_{\del\xi}^2-2\VEV{n^2}_{\del\xi}\VEV{n}_{\del\xi}^2
+\VEV{n}_{\del\xi}^4-\VEV{n}_{\del\xi}^2]^{-1}
$.

Notice that $P_{\del\xi}(n)$ is expressed as $\gamma^n$-weighted
cluster $l$-sum of $n$-particle
productions; which can also be reexpressed as a product of Poisson
distribution in cluster number $l$ and
negative binomial in $(n-l)$:
$P_{\del\xi}(n) =\Sigma_{l=1}^n ~\{[\gamma c/(1-\gamma)]^l/l!~exp[-\gamma
c/(1-\gamma)]\}
\{(1-\gamma)^l \gamma^{n-l} C^{n-1}_{l-1} \}$.
Interestingly such distributions are also those given by the
Poisson-distributed cluster models
\attach{[13]}.

Next we determine from data $\VEV{n}_{\del\xi}$ and $\VEV{n^2}_{\del\xi}$,
thus $A_{\del\xi}$ and $B_{\del\xi}$ in Eq.(3).
Figs.(1a) and (1b) show the values
of $A_{\del\xi}$ and $B_{\del\xi}$ at various values of $\del\xi$ obtained
from ${\bar p}p$ data\attach{[5]}.
For $e^+e^-$ and $NN$ reactions we can not make such detail analysis
since only $\VEV{n}_{\Delta\xi}$
and $\VEV{n^2}_{\Delta\xi}$ are available, shown in Fig.(2); thus
we can only obtain $A_{\Delta\xi}$ and $B_{\Delta\xi}$, shown in
Fig.(1c) for $e^+e^-$.
Fortunately, for analysing universality $\F_q[F_2]$, we need only data
of $F_q(\del\xi)$ and $F_2(\del\xi)$, not those of $\VEV{n}_{\del\xi}$ and
$\VEV{n^2}_{\del\xi}$.

\noindent{\bf Predictions, Comparisons, and Outlook}

Once the two parameters $A_{\del\xi}$ and $B_{\del\xi}$ are fixed
via $\VEV{n}_{\del\xi}$ and $\VEV{n^2}_{\del\xi}$, Eqs.(4) and (5) specify
completely the multiparticle
distributions from the 1-d Ising model; thus they constitute as predictions
of the model.  The $n$-distributions of
$P_{\del\xi}(n)$ given in Eq.(4) for various values of $\del\xi$ are shown in
Fig.(1c)
together with data from
${\bar p}p$ reaction at $\sqrt{s}=200$ GeV and $\del\xi=$ 3, 6, and 10; (at
$\sqrt{s}=540$ and $900$ GeV,
the results are similar; not shown).  The results are good, better for larger
$\del\xi$.
\attach{[14]}
Also shown in Fig.(1c) are the $n$-distributions of $P_{\del\xi}(n)$
at $\del\xi =\Delta\xi$ together with
determined $A_{\Delta\xi}$ and $B_{\Delta\xi}$ for $e^+e^-$ reactions at
$\sqrt{s}=$ 91 GeV, $\Delta\xi$=10; and at $\sqrt{s}=$ 34 GeV, $\Delta\xi$=10.

It is interesting
to note in Figs.(1a), (1b) that $A_{\del\xi}$ and $B_{\del\xi}$ determined
from $\bar{p}p$ data flaten out in the larger
$\del\xi$ region, $\del\xi \geq 3$,
which means that for $\del\xi \geq 3$ the Ising model with
$\del\xi$-independent coupling constants actually can correctly describe
the $\del\xi$-distribution
from its phase-space factor $M=\Delta\xi/\del\xi$ alone.
This feature is important to be checked out in future higher-energy data where
larger $\del\xi$ intervals will become available.

We note from Eq.(4) that Ising $P_{\Delta\xi}(n)$ has no KNO scaling,
\attach{[8]} \ie~~
$[\VEV{n}_{\Delta\xi} P_{\Delta\xi}(n)]$ is not a scaling function of
$[n/ \VEV{n}_{\Delta\xi}]$ though approximate KNO scaling can be
true for some limited energy-range.
Also clear is that the function forms of our Ising $P_{\del\xi}
(n)$ in Eqs.(4) and (5) are different from those of single-negative-binomial
\attach{[15]}, though numerical comparison
at present energies are quite similar,
see solid and dashed lines in Fig.(1c).

{}From the $P_{\delta\xi}(n)$'s of Eq.(4) we can calculate $F_q(\delta\xi)$.
Due to the $\gamma^n$-dependence of the $\{~\}$-part in $P_{\del\xi}(n)$
and the nice expression of $z_p$ from Ising, Eq.(5),
we obtain the following neat results
$$
F_q(\delta\xi) = [{\gamma^q \over z_p}
{\partial^q z_p \over (\partial \gamma)^q}]
[{\gamma \over z_p}{\partial z_p \over \partial \gamma}]^{-q}
{}~,\eqno(6)
$$
which in turn give
$$
\F_q[F_2]  = 1
+\Sigma_{l=1}^{q-1} {(q-1)! q! \over l! (q-l-1)! (q-l)! 2^l}
[F_2(\del\xi) -1]^{l} ~~,\eqno(7.a)
$$
where
$$
[F_2(\del\xi)-1] =
2 {1-e^{-2B_{\del\xi}}\over A_{\del\xi}} {\Delta\xi\over \del\xi} ~~.\eqno(7.b)
$$
Eqs.(7) indicate that all the dependences on the
three parameters of the model, $A_{\del\xi}, B_{\del\xi}$, and
$\del\xi/\Delta\xi$, are absorbed in $F_2$; thus $\F_q[F_2], q=3,4,5,\cdots$,
are universal functions, depending solely on $F_2$.

Since $\VEV{n(n-1)\cdots(n-q+1)}_{\del\xi}$ is also the fully-integrated
$q$-particle
inclusive crossection in the interval $\del\xi$, the universality
$\F_q[F_2]$ implies that all fully-integrated multiparticle
inclusive crossections are functions of fully-integrated 2-particle inclusive
crossection in the interval $\del\xi$.
This is an essential feature captured by the next-neighbor agitation in the
Ising model.

The scaling and universality of Eqs.(6) and (7) were obtained by us in Ref.[1]
from a different calculation
bypassing the calculation of $P_{\del\xi}(n)$.
In the present calculation, we can identify
the characteristics in
$P_{\del\xi}(n)$ that lead to scaling and universality in $F_q$:
$\gamma^n$-dependence in $P_{\del\xi}(n)$, and
${\partial z_p / \partial \gamma}= g~z_p$, where
$g$ has the property that
$[g^{-3}\partial^2g/(\partial \gamma)^2]$ is a function of
$[g^{-2}\partial g/\partial \gamma]$.
Ising model provides such an explicit
$g=c/(1-\gamma)^2$ and
$
[g^{-3}\partial^2g/(\partial \gamma)^2]=(3/2)
[g^{-2}\partial g/\partial \gamma]^2
$.
Single-negative-binomial distribution for $P_{\del\xi}(n)$ has similar
properties, (which we discuss in Ref.[15]~)
except its $g$ satisfies an equation with 3/2 on the right-hand-side replaced
by 2.

The universal curve of $\F_q[F_2]$
given in Eq.(7) for $q=3$ is plotted in
solid lines in Fig.(3a).
The agreement with the data is suprisingly good (excellent for smaller
$F_2(\del\xi)$ region),
considering they being parameter-free predictions.
(Similar good results have been obtained for $q=4$ and $5$ for all the
listed reactions\attach{[5,6,7]}; shown in Figs.(4a) and (5a)).

{}From data-points of $F_2(\del\xi)$ in Fig.(2)
we can also obtain $F_q(\del\xi), q=3,4,5,\cdots$.
They are shown as solid
lines in Fig.(3b) for $q=3$ and compared to data from Refs. [5,6,7]. (Similar
results have been obtained for
$q=4$ and $5$ for all listed reactions\attach{[5,6,7]};
shown in Figs.(4b) and (5b)).
Those given by single-negative-binomials are almost exactly the same as ours
for
$F_2 \rightarrow 1$, and better for $F_2 > 1.5$~.

Finally we make a few concluding remarks.
We have identified the salient features in $P_{\del\xi}(n)$ that lead to
scaling and universality and have provided
a framework to analyze data of high-energy
multiparticle productions in $pp$, $\bar{p}p$, $ep$, $e^+e^-$, and $NN$.
The important future tests are in the
larger $\del\xi$ regions.
If the trend of constancy of $A_{\del\xi}$ and $B_{\del\xi}$ continues as
the available $\del\xi$-interval increases at higher-energies, the
$\del\xi$-dependence
in $\VEV{n}_{\del\xi}, \VEV{n^2}_{\del\xi}$, and $F_q(\del\xi)$ can all be
anticipated :
from Eq.(3), $\VEV{n}_{\del\xi} \sim \del\xi$,
$\VEV{n^2}_{\del\xi} \sim (\del\xi)^2$;
from Eq.(7),
$[F_2(\del\xi)-1]\sim (\del\xi)^{-1}$ and all $F_q(\del\xi)\rightarrow 1$.
They are shown as dash-dot lines
in Figs.(2), (3b), (4b) and (5b) for the $\bar{p} p$ reaction.
Notice that they are straight lines in Fig.(2) showing $ln[F_2(\del\xi)-1]$
vs $ln(\del\xi)$.
As $F_2$ decreases for larger $\del\xi$, there will also be more region
available in
$F_2$ to check universality $\F_q[F_2]$.
We hope that soon $\VEV{n}_{\del\xi}$, $\VEV{n^2}_{\del\xi}$, and
$P_{\del\xi}(n)$ will also be provided from other reactions besides
$\bar{p} p$, so full analysis as shown in Fig.(1) can also be done.
It will be
interesting to see if the gluon jets will have the same universal
$\F_q[F_2]$ as the quark jets in high-energy nuclear reactions.
We need measurements of $F_q$ in the 1-d variable $\del\eta$ in high-energy
nuclear reactions,
rather then the current available data in $\del\eta\cdot\del\phi$.\attach{[7]}
These are important measurements for
future relativistic heavy-ion experiments.

Indeed, we look forward to more higher-energy multiparticle-production data
in different reactions to be further studied and compared with these
salient features learned from 1-d Ising model.

\vfil

\noindent{\bf Acknowledgements}

We thank professor P. Carruthers and professor I. Sarcevic for informative
discussions; professor Y.-F. Cheng, professor Y.-C. Lin,
and professor C.-T. Lin,
organizers of the Six$^{\rm th}$ Spring School on Particle Physics,
at National Central University, Taiwan, April 2-7, 1992,
for a lively conference where we had obtained much information about
the current status of the data.

The work is partly supported by the US Department of Energy (DOE)
and INCOR of University of California.

\vfill

\centerline{\bf Figure Captions}
\parindent -25pt

Figs.(1a), (1b){\phantom{...}}$A_{\del\xi}$ and $B_{\del\xi}$ of Eq.(3)
determined from data $\VEV{n}_{\del\xi}$ and $\VEV{n^2}_{\del\xi}$ of
$\bar{p} p$, Ref.~[5].
Notice that at fixed $\del\xi=\Delta\xi$, $B_{\Delta\xi}$ increases with
energy, so $\langle n^2\rangle_{\Delta\xi}-\langle n\rangle^2_{\Delta\xi}$
from Eq.(3) increases faster than $\langle n\rangle_{\Delta\xi}$ in energy.

Fig.(1c){\phantom{...}}$P_{\del\xi}(n)$ as function of $n$.
(Results for $\bar{p}p$ at $\sqrt{s}$=540 and 900 GeV are as good; not shown)

Fig.(2){\phantom{...}}$[F_2(\del\xi)-1]$ as function of $\del\xi$.
The dash-dot lines, which are straight lines in this $ln [F_2(\del\xi)-1]$
vs $ln(\del\xi)$ plot, are anticipated Ising results for $\del\xi > 10$
assuming that $A_{\del\xi} , B_{\del\xi}$
will stay constant at their present values at $\del\xi =10$ as shown
in Figs.(1a) and (1b).

Fig.(3a){\phantom{...}}Universal curve of $\F_3[F_2]$ as function of $F_2$.

Fig.(3b){\phantom{...}}$F_3(\del\xi)$ from the same data as indicated
in Fig.(2) for $F_2(\del\xi)$.
The dash-dot lines are anticipated Ising results for $\del\xi > 10$ assuming
that $A_{\del\xi} , B_{\del\xi}$
will stay constant at their values at $\del\xi =10$ as shown in
Figs.(1a) and (1b).

Figs.(4a), (5a){\phantom{...}}Same as Fig.(3a), except for $\F_4[F_2],
\F_5[F_2]$, respectively.

Figs.(4b), (5b){\phantom{...}}Same as Fig.(3b), except for $F_4(\del\xi),
F_5(\del\xi)$, respectively.

\vfill
\endpage

\hoffset=-0.3in
\hsize=7.1in
\voffset=-0.25in
\vsize=9.4in

\centerline{\bf References}

\pointbegin
L. L. Chau and D. W. Huang, Phys. Lett. \underbar{B283} (1992) 1.

\point
A. Bia{\l}as and R. Peschanski, Nucl. Phys. \underbar{B273} (1986) 703;
\underbar{B308} (1988) 857.

\point
P. Carruthers and C. C. Shih, Int. J. Mod. Phys. \underbar{A2} (1987) 1447;
P. Carruthers and I. Sarcevic, Phys. Rev. Lett. \underbar{63} (1989) 1562;
P. Carruthers, Int. J. Mod. Phys. \underbar{A4} (1989) 5587;
P. Carruthers, H. C. Eggers and I. Sarcevic, Phys. Rev. \underbar{C44}
(1991) 1629; Phys. Lett. \underbar{B254} (1991) 258.

\point
W. Ochs and J. Wosiek, Phys. Lett. \underbar{B214} (1988) 617,
\underbar{B232} (1989) 271;
W. Ochs, Z. Phys. \underbar{C50} (1991) 339;
and in {\it Fluctuations and Fractal Structure}, eds. R. C. Hwa {\it et al.},
World Scientific, 1991.

\point
$\bar p p$ experimental data are from CERN experiment UA5, Phys. Rep.
\underbar{154} (1987) 247; Z. Phys. \underbar{C43} (1989) 357.

\point
$e^+e^-$ experimental data are from CERN experiment ALEPH, Z. Phys.
\underbar{C53} (1992) 21.

\point
The nuclear experimental data are from the KLM Collaboration, R. Holynski {\it
et al.}, Phys. Rev. \underbar{C40} (1989) 2449.
Strictly speaking, our 1-d Ising results are not applicable to the nuclear 2-d
plot in $\del\eta\cdot\del\phi$.  We include here to entise future true 1-d
nuclear data in $\del\eta$ with good statistics.

\point
Z. Koba, H. B. Nielsen and P. Olesen, Nucl. Phys. \underbar{B40} (1972) 317.

\point
M. Plank, Sitzungsber. Deutsch. Akad. Wiss. Berlin \underbar{33} (1923) 355;
P. K MacKeown and A. W. Wolfendale, Proc. Phys. Soc. \underbar{89} (1966) 553;
A. Giovannini, Nuovo Cimento \underbar{A15} (1973) 543; \underbar{A24} (1974)
421;
N. Suzuki, Progr. Theor. Phys. \underbar{51} (1974) 1629;
W. J. Knox, Phys. Rev. \underbar{D10} (1974) 65;
M. Garetto {\it et al.}, Nuovo Cimento \underbar{A38} (1977) 38;
P. Carruthers and C. C. Shih, Phys. Lett. \underbar{B127} (1983) 242.

\point
A. Giovannini and L. Van Hove, Z. Phys. \underbar{C30} (1986) 391;
Acta Phys. Pol. \underbar{B19} (1988) 917, 931.

\point
CERN experiment
UA5, R.E. Ansorge {\it et al.}, Z. Phys. \underbar{C43} (1989) 357.

\point
Note that if for $A_{\del\xi}$ and $B_{\del\xi}$ constant in $\del\xi$,
especially in the small $\del\xi$ regions, Ising results do give intermittent
behavior $F_q(\del\xi)\rightarrow (1/\del\xi)^{q-1}$  as $\del\xi\rightarrow
0$, see Eq.(7).
In Ref.[1] we tried to obtain the $\del\xi$ dependence by interpreting
$M=(\Delta\xi/\del\xi)^{1/\alpha}$ and fitted data to
find out the value of $\alpha\sim 5$ to flatten the $\del\xi$-dependence
of $F_q(\del\xi)$.  Here we have changed the strategy.  Rather than
presuppose that all the $\del\xi$-dependence is in $M$, we leave it open and
find
out from data what are the dependence on $\del\xi$ in the two parameters of
the model $A_{\del\xi}$ and $B_{\del\xi}$ when we take the most natural
interpretation $M=(\Delta\xi/\del\xi)$.
In doing so the $\alpha\ne 1$ effect is
reflected in that $A_{\del\xi}$ and $B_{\del\xi}$
having rather strong dependence on $\del\xi$ in the small $\del\xi$ region,
see Figs.(1a) and (1b).
In any case, the
existence of intermittency in multiparticle-particle-production is hard
to establish unambiguously due to the following intrinsic difficulties: the
error
in data increases
as $\del \xi$ decreases; as $\del \xi$ decreases to a value such
that the average number of particles produced in that width of $\del \xi$
become comparable or smaller than $q$, \ie~ $\VEV{n}_{\Delta\xi}\cdot
{(\del \xi/\Delta \xi)} \lsim q$, the factorial moments
$F_q(\del \xi)$ is no more a good quantity to measure fractal behavior.
(This is the same situation as measuring coast line with a ruler
smaller than the wavelength of the waves at the shore.)  Such
difficulties will not be overcome even if the energies of reactions increase
since that will only increase the range of rapidity but not improving the
errors in small-$\del \xi$ measurements of $F_q$ nor increasing the number of
particles in given $\del\xi$.  That is why there are
many different opinions on intermittence studies;\attach{[1,16]} thus here we
focus and emphasize the larger $\del\xi$ region.

\point
J. Finkelstein, Phys. Rev. \underbar{D37} (1988) 2446.
This work gives yet another physical interpretation to our 1-d Ising
$P_{\del\xi}(n)$.
We thank Dr. Finkelstein for informing us about this point and his paper.

\point
It is forseeable that the agreement of the 1-d Ising results with data
can be improved by fine-tuning the model, \eg~ including the clustering
effects; varing $\varepsilon$
to include site-dependence in $\varepsilon$; or introducing
next-nearest neighbor interactions.

\point
The negative binomial distribution\attach{[9,10,11]} is given by
$$
P_{\del\xi}^{NB}(n)
=(1/z_p^{NB})\{ (\gamma^{NB})^n  [(n+k_{\del\xi}-1)!]
[ n! (k_{\del\xi}-1)! ]^{-1} \},
$$
where $\gamma^{NB}=\VEV{n}_{\del\xi}/(k_{\del\xi}+\VEV{n}_{\del\xi})$,
$(k_{\del\xi})^{-1}=\VEV{n^2}_{\del\xi}/\VEV{n}_{\del\xi}^2
-1/\VEV{n}_{\del\xi} -1$
and $z_p^{NB}=(1-\gamma^{NB})^{-k_{\del\xi}}$,
where $\VEV{n}_{\del\xi}$ and $\VEV{n^2}_{\del\xi}$ are supposed to be
determined from experiments.
Our Ising formulation has exactly the same degrees of freedom
determined from data, but the dependences of $P_{\del\xi}(n)$ on
$\VEV{n}_{\del\xi}$ and $\VEV{n^2}_{\del\xi}$ are very different.
Following similar discussions as for the Ising case,
one can easily derive the factorial moments
$F_q^{NB}(\del\xi)$ using a equation exactly like Eq.(6) and obtain the
explicit expressions for universality
$\F_q^{NB}[F_2^{NB}]$.
This universality in $NB$ can be traced to the specific form of
$z_p^{NB}$ which leads a equation similar to Eq.(8), see the discussions
in the text after Eq.(8).

\point
H. Satz, Nucl. Phys. \underbar{B326} (1989) 613;
R. Peschanski, Nucl. Phys. \underbar{B327} (1989) 144;
P. Dahlqvist, B. Andersson and G. Gustafson, Nucl. Phys. \underbar{B328} (1989)
76;
A. Capella, K. Fialkowski and A. Krzywicki, Phys. Lett. \underbar{B230} (1989)
149;
J. Dias de Deus and J. C. Seixas, Phys. Lett. \underbar{B229} (1989) 402,
\underbar{B246} (1990) 506;
X. N. Wang, Phys. Lett. \underbar{B248} (1990) 447.

\vfill
\end